\documentclass{article}

\usepackage[english]{babel}
\usepackage{amsmath}
\usepackage{amsthm}
\usepackage{amsfonts}

\usepackage{lmodern}
\usepackage{tikz}
\usetikzlibrary{arrows}
\usepackage{cite}
\usepackage{hyperref}

\newtheorem{defi}{Definition}
\newtheorem{thm}{Theorem}

\newcommand{\etal}{\textit{et al. }}
\newcommand{\ie}{\textit{i.e. }}
\newcommand{\enc}{\mathrm{Enc}}
\newcommand{\dec}{\mathrm{Dec}}
\newcommand{\tfs}{\mathrm{Tamper}^f_s}
\newcommand{\mf}{\mathcal{F}}
\newcommand{\md}{\mathcal{D}}
\newcommand{\bit}{\{0,1\}}
\newcommand{\same}{\mathbf{same}}
\newcommand{\ff}{\mathbb{F}}
\begin{document}

\title{Non-Malleable Codes from the Wire-Tap Channel\thanks{This work has been partially funded by the ANR SPACES project.}}

\author{
Herv\'e Chabanne \thanks{IDentity \& Security Alliance (The Morpho and T\'el\'ecom ParisTech Research Center)
T\'el\'ecom ParisTech --
46, rue Barrault - 75013 Paris - France --
Email: \{chabanne, cohen, flori, patey\}@telecom-paristech.fr}
\thanks{Morpho --
11, boulevard Gallieni - 92130 Issy-Les-Moulineaux - France --
Email: \{herve.chabanne, alain.patey\}@morpho.com}
\and G\'erard Cohen \footnotemark[2] \thanks{CNRS-LTCI}
\and Jean-Pierre Flori \footnotemark[2] \footnotemark[4]
\and Alain Patey \footnotemark[2] \footnotemark[3] \footnotemark[4]
}

\tikzstyle{carre}=[draw, minimum size=2em]
	
\maketitle

\begin{abstract}

Recently, Dziembowski \etal introduced the notion of \emph{non-malleable codes} (NMC), inspired from the notion of non-malleability in cryptography and the work of Gennaro \etal in 2004 on tamper proof security. Informally, when using NMC, if an attacker modifies a codeword, decoding this modified codeword will return either the original message or a completely unrelated value.

The definition of NMC is related to a family of modifications authorized to the attacker. In their paper, Dziembowski \etal propose a construction valid for the family of all bit-wise independent functions.

In this article, we study the link between the second version of the Wire-Tap (WT) Channel, introduced by Ozarow and Wyner in 1984, and NMC. Using coset-coding, we describe a new construction for NMC w.r.t. a subset of the family of bit-wise independent functions. Our scheme is easier to build and more efficient than the one proposed by Dziembowski \etal

\end{abstract}

\section{Introduction}

In cryptography, the non-malleability property  \cite{DDN91} requires that it is impossible, given a ciphertext, to produce another different ciphertext so that the corresponding plaintexts are related to each other. Non-malleability under adaptive chosen-ciphertext attack (NM-CCA2) is one of the strongest computational security property that is required from an asymmetric encryption scheme (it is equivalent to indistinguishability under adaptive chosen-ciphertext attack (IND-CCA2)).

Recently, Dziembowski \etal \cite{DPW10} proposed a transposition of the cryptographic definition of non-malleability to the field of coding theory. 
Informally, they define a NMC as a code such that, when a codeword is subject to modifications, its decoding procedure either corrects these errors and decodes to the original message or returns a value that is completely unrelated to the original message.

The property of non-malleability, as defined in \cite{DPW10}, is subject to a choice of a family of modifications that we allow an adversary to make on the codewords. Dziembowski \etal also proved that it is impossible for a code to be non-malleable w.r.t. the set of all possible modifications of codewords.

The motivation for NMC is tamperproofness. The authors of \cite{DPW10} were indeed much influenced by the work of Gennaro \etal \cite{GLMMR04}. Non-malleability can be useful in real-life applications. Some storage devices may be assumed to be ``read-proof'' because of a sufficient amount of physical or algorithmic protections to prevent anyone from learning the data stored on them. However, even if one cannot read the data, injecting faults in the data and observing the way it affects functions using these data can help to recover them. Injecting faults can be done for instance using lasers \cite{SA02}. There exists an important literature on how to use Differential Fault Analysis to break cryptosystems (e.g. \cite{BDL97,BCNTW04}).

Dziembowski \etal studied deeply the non-malleability w.r.t. bit-wise independent tampering functions, \ie modifications that affect each bit of the codeword independently: flipping the bit or setting it to 0 or 1. This is typically what can be done using fault injections and, consequently, focusing on this family of tampering functions is worthwhile.

In \cite{DPW10}, a construction for NMC w.r.t. all bit-wise independent functions is proposed. However, an implementable construction is left as an open problem. Our goal is to propose NMC that can be explicitly built. To this end, we exploit a relation that can be established between the model for NMC and the second version of the Wire-Tap channel \cite{OW84}. This allows us to prove how coset-coding can be used to build a NMC. Furthermore, the decoding procedure of linear-coset coding consists uniquely of one matrix-vector product. Our construction is thus computationally efficient. Moreover, unlike their solution, our procedure always decodes messages whereas theirs is closer to error detection and often returns an error symbol.

\subsection*{Organization of the Paper}

In Section~\ref{sec:nmc}, we explain and give the formal definitions for NMC as established in \cite{DPW10}. We describe the model of the WT channel in Section~\ref{sec:wire} and explain the use of coset-coding. We show how the second version of the WT channel and NMC w.r.t. bit-wise independent functions are related and prove why coset-coding can be used as a NMC in Section~\ref{sec:scheme}. We finally conclude in Section~\ref{sec:conclu}.

\section{Non-Malleable Codes}
\label{sec:nmc}

In this section, we intend to give an easy-to-understand description of NMC and their goals. All definitions come from \cite{DPW10}.

In the following, we consider a randomized encoding function $\enc: \bit^k \mapsto \bit^n$, which is associated to a deterministic decoding function $\dec: \bit^n \mapsto \bit^k \cup \{\bot\}$, where $\bot$ means that the codeword cannot be decoded. Let $\ff_2$ denote the field with two elements.

\subsection{The Tampering Experiment}
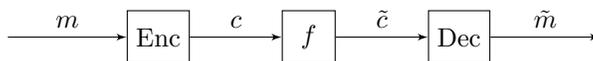
\begin{figure}[h!]
\begin{center}

\begin{tikzpicture}[node distance=2cm,auto,>=latex']
    \node [carre] (a) {Enc};
    \node (begin) [left of=a,node distance=2cm, coordinate] {};
    \node [carre] (b) [right of=a] {$f$};
    \node [carre] (c) [right of=b] {Dec};
    \node (end) [right of=c, node distance=2cm]{};
    \path[->] (begin) edge node {$m$} (a);
    \path[->] (a) edge node {$c$} (b);
    \path[->] (b) edge node {$\tilde{c}$} (c) ;
    \path[->] (c) edge node {$\tilde{m}$} (end); 
    
\end{tikzpicture}

\caption{The Tampering Experiment}
\label{fig:exp}

\end{center}

\end{figure}

Let us first introduce the situation considered in NMC. In this model, a source message $m$ is encoded using $\enc$, in order to be later decoded using $\dec$. The codeword $c = \enc(m)$ is stored on a device or sent over a channel before being decoded. During this phase, an attacker applies some tampering function $f$ belonging to a given family of functions $\mf \subset {\mathbb{F}_2^n}^{\mathbb{F}_2^n} $ . A tampered codeword $\tilde{c} = f(c)$ is thus obtained. This erroneous codeword is then decoded to $\tilde{m}=\dec(\tilde{c})$. This process is described in Figure~\ref{fig:exp}.

Now focus on the behaviour of the attacker, called Eve in the following. Eve applies a function $f \in \mf$ to the codeword $c$, but she does not read $c$. In the real world, this can be seen as injecting faults on a device that you cannot read (e.g. a smart-card) using, for instance, a laser. In this experiment, Eve can however read the resulting decoded message $\tilde{m}$ and try to learn as much as possible about $m$ from $\tilde{m}$. Let us also specify that $f$ is a deterministic function and, furthermore, that Eve knows which function she has chosen in $\mf$.

\subsection{Defining Non-Malleability}

Let us now give the formal definition of non-malleability. Let $\mf$ be a family of tampering functions. For each $f \in \mf$, we define a random variable $\tfs$ corresponding to the tampering experiment described in the previous section:

\[
\tfs = \left\{ \begin{array}{c} c \leftarrow_R \enc(s), \tilde{c} = f(c) , \tilde{s} = \dec(\tilde{c}) \\
 \mathrm{Output}: \tilde{s}
   \end{array} \right\}
\]

The randomness is induced by the encoding function $\enc$.

The \emph{Non-Malleability} property is defined as follows:

\begin{defi}[Non-Malleability]
Let $(\enc,\dec)$ be a coding scheme, where $\enc: \bit^k \mapsto \bit^n$ is random and  $\dec: \bit^n \mapsto \bit^k \cup \{\bot\}$ deterministic. Let $\mf \subset {\ff_2^n}^{\ff_2^n}$ be a family of tampering functions.

We say that the coding scheme $(\enc,\dec)$ is \emph{non-malleable w.r.t. $\mf$} if for each $f \in \mf$, there exists a distribution $\md_f$ over $\bit^k \cup \{\bot,\same\}$ such that, $\forall s \in \bit^k$, we have:

\begin{equation}\label{eq:tamper}
\tfs \approx \left\{ \begin{array}{c}  \tilde{s} \leftarrow \md_f \\
 \mathrm{Output}
 \left\{ \begin{array}{c}  s \ \mathrm{if} \ \tilde{s} = \same \\ \tilde{s} \ \mathrm{otherwise}\end{array}\right.
   \end{array} \right\}
\end{equation}
where $\approx$ denotes computational or statistical indistinguishability.

\end{defi}

\subsection{Explaining the Definition}

First, notice that the definition is relative to a family $\mathcal{F}$ of tampering functions, but the property of indistinguishability concerns each function $f$ separately. Non-malleability w.r.t. a family is in fact non-malleability w.r.t. each function in this family.

Now let us recall what we expect from a NMC. We want that, after the tampering experiment, either the codeword $\tilde{c}$ is well-decoded to the original message $s$ despite the tampering or the decoding procedure results in a value $\tilde{s}$ that is unrelated to the original message. That is the idea behind the distribution $\md_f$: either it returns the symbol $\same$, meaning that the decoding furnishes the original value or it returns a value $\tilde{s} \in \bit^k \cup \{\bot\}$. As $\md_f$ depends only on $f$ and not on the message $s$, in the latter case, the value returned in the second part of Equation~(\ref{eq:tamper}) is unrelated to $s$.

\subsection{Basic Examples}

We summarize here two examples developed in \cite{DPW10} that correspond to usual families of codes encompassed by the definition of NMC.

\subsubsection*{Error Correction}

Let us assume that $\mf$ is a family of tampering functions and $C$ an error-correcting code such that errors introduced by the application of a function $f \in \mf$ on any codeword of $C$ can be corrected. Then $C$  is non-malleable w.r.t. $\mf$. The distribution associated to every function $f \in \mf$ is the constant distribution $\md_f = \same$, since erroneous codewords are always well-decoded.


\subsubsection*{Error Detection}

The same idea can be applied to error-detecting codes. If there is a family $\mf$ of tampering functions such that each $f \in \mf$ introduces errors in every codeword that are detected by a code $C$, then $C$ is non-malleable w.r.t. $\mf$. The distribution associated to every function $f \in \mf$ is the constant distribution $\md_f = \bot$.

\subsection{General (Im)Possibility Results}

\subsubsection*{Impossibility}
As proven in \cite{DPW10}, no code is non-malleable w.r.t. the set of all possible tampering functions (\ie $\mf = {\ff_2^n}^{\ff_2^n}$). Indeed there is, for instance, in $\mf$ a function that decodes the codeword, ``increments''  the message (\ie adds 1 to its representation in $\ff_2^k$) and re-encodes it. The result of the decoding of such a tampered codeword would always be $s+1$ and thus would be neither the original message $s$ nor an unrelated value.

\subsubsection*{Possibility}
In \cite{DPW10}, the authors prove that for any bounded-sized family of tampering functions, there exists a NMC. Their result is summed up in the following theorem:

\begin{thm}[\cite{DPW10}]

Let $\mf \subset {\ff_2^n}^{\ff_2^n}$ be a family of tampering functions such that $n > \log(\log(|\mf|))$. Then there exists a non-malleable code w.r.t. $\mf$.

\end{thm}

\subsection{Bit-wise Independent Tampering}

Bit-wise independent tampering is a special case of tampering where each bit of the codeword is tampered with independently. Formally a function $f: \bit^n \mapsto \bit^n$ is bit-wise independent if we can find $n$ independent functions $f_1,\ldots,f_n : \bit \mapsto \bit$ such that $\forall x \in \bit^n, f(x)=(f_1(x),\ldots,f_n(x))$. There are four possibilities for each $f_i$ which we denote by \textbf{keep}, \textbf{flip}, \textbf{0} and \textbf{1} (\textbf{keep} and \textbf{flip} are explicit, \textbf{0} (resp. \textbf{1}) is the function that sets a bit to 0  (resp. 1) regardless of what it was before).

In \cite{DPW10}, a construction for a NMC w.r.t. the family of all bit-wise independent functions is introduced. It uses Linear Error-Correcting Secret-Sharing (LECSS) schemes \cite{CCGHV07} and Algebraic Manipulation Detection (AMD) codes \cite{CDFPW08}. Both are quite new tools and even the authors of \cite{DPW10} leave the explicit construction of LECSS codes as an ``interesting open problem''. Furthermore, their solution is quite close to error detecting codes as it decodes to $\bot$ after a tampering in most cases\footnote{In their proof of non-malleability, the authors of \cite{DPW10} distinguish different cases depending on the considered tampering function (more precisely its number $q$ of $\mathbf{0}$ and $\mathbf{1}$ sub-functions) and the \emph{secrecy} $t$ of the LECSS scheme. When $t < q < n-t$, the tampering experiment always returns $\bot$ and when $q \le t$, the scheme is likely to often return $\bot$.}.

In Section~\ref{sec:scheme}, we propose a new way to build NMC w.r.t. bit-wise independent functions. Our solution covers less tampering functions but uses more standard and efficient tools. Moreover, our scheme is neither error-correcting nor error-detecting (it never returns $\bot$) and so, to our opinion, is closer to the original definition of non-malleability, which is more generic than error detection or correction.

\section{The Wire-Tap Channel}
\label{sec:wire}

In the following, a \emph{$[n,k,d]$ linear code} denotes a subspace of dimension $k$ of $\ff_2^n$ with minimal Hamming distance $d$.

\subsection{Linear Coset Coding}

\begin{figure}[h!]
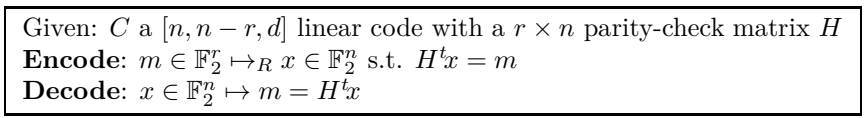


\centering
\fbox{
\begin{minipage}{0.9\linewidth}
Given: $C$ a $[n,n-r,d]$ linear code with a $r \times n$ parity-check matrix $H$\\
\textbf{Encode}: $m \in \mathbb{F}_2^r \mapsto_R x \in \mathbb{F}_2^n$ s.t. $H ^t \! x = m$ \\ 
\textbf{Decode}: $x \in \mathbb{F}_2^n \mapsto m=H ^t \! x$
\end{minipage}
}
\caption{Linear Coset-coding}
\label{fig:coset}
\end{figure}

Coset coding is a random encoding used for both models of WT Channel. This type of encoding uses a $[n,k,d]$ linear code $C$ with a parity-check matrix $H$. Let $r = n-k$. To encode a message $m \in \mathbb{F}_2^r$, one chooses randomly an element among all $x \in \mathbb{F}_2^n$ such that $m =H ^t \! x$. To decode a codeword $x$, one just applies the parity-check matrix $H$ and obtains the syndrome of $x$ for the code $C$, which is the message $m$. This procedure is summed up in Figure~\ref{fig:coset}.

\subsection{The Wire-Tap Channel I}

\begin{figure}[h!]
\begin{center}

\begin{tikzpicture}[node distance=2cm,auto,>=latex']
  \draw
    (0,0) node [] (a) {Alice}
    node [right of=a, carre, node distance=1.8cm] (b) {Enc}
    node [right of=b, coordinate, node distance=.7cm] (begintap) {}
    node [right of=b, node distance=3cm,carre, draw] (c) {small (or no) noise}
    node [below of=c, node distance=2cm,carre] (endtap) {big noise}
    node [right of=c,node distance=3cm] (d) {Bob}
    node [right of=endtap,node distance=3cm](e){Eve};

  \path[->] (a) edge node {$m$} (b);
  \path[->] (b) edge node {$c$} (c);
  \path[->] (c) edge node {$c'$} (d);
  \path[->] (begintap) edge node {} (endtap);
  \path[->](endtap)edge node {$c''$}(e);
\end{tikzpicture}
\caption{The Wire-Tap Channel I}
\label{fig:wire1}
\end{center}
\end{figure}
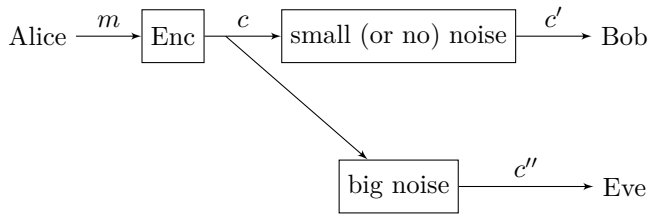

The Wire-Tap Channel was introduced by Wyner \cite{Wyn75}. In this model, a sender Alice sends messages over a potentially noisy channel to a receiver Bob. An adversary Eve listens to an auxiliary channel, the WT channel, which is a noisier version of the main channel. It was shown that, with an appropriate coding scheme, the secret message can be conveyed in such a way that Bob has complete knowledge of the secret and Eve does not learn anything. In the special case where the main channel is noiseless, the secrecy capacity can be achieved through a linear coset coding scheme. We summarize the WT Chanel I in Figure~\ref{fig:wire1}.

\subsection{The Wire-Tap Channel II}

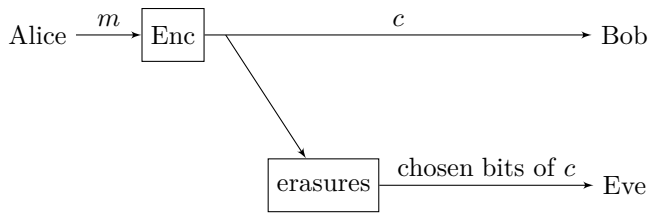
\begin{figure}[h!]
\begin{center}

\begin{tikzpicture}[transform shape,node distance=2cm,auto,>=latex']
  \draw
    (0,0) node [] (a) {Alice}
    node [right of=a, carre, node distance=1.8cm] (b) {Enc}
    node [right of=b, coordinate, node distance=.7cm] (begintap) {}
    node[right of = b, coordinate, node distance=2 cm](c){}
    node [below of=c, node distance=2cm, carre] (endtap) {erasures}
    node [right of=b,node distance=6cm] (d) {Bob}
    node [below of=d,node distance=2cm](e){Eve};

  \path[->] (a) edge node {$m$} (b);
  \path[->] (b) edge node {$c$} (d);
  \path[->] (begintap) edge node {} (endtap);
  \path[->](endtap)edge node {chosen bits of $c$}(e);
\end{tikzpicture}

\caption{The Wire-Tap Channel II}
\label{fig:wire2}
\end{center}
\end{figure}

Ten years later, Ozarow and Wyner introduced a second version of the WT Channel \cite{OW84}. In this model, both main and WT channels are noiseless. This time, the disadvantage for Eve is that she can only see messages with erasures: she has only access to a limited number of bits per codeword. She is however allowed to choose which bits she can learn. We summarize the Wire-Tap Chanel II in Figure~\ref{fig:wire2}.

The encoding used in this model is again a coset coding based on a linear code $C$, as in the Wire Tap Channel I with a noiseless main channel. Let $d^{\bot}$ denote the minimal distance of the dual $C^{\bot}$ of $C$. One can prove (see \cite{Wei91} for instance) that, if Eve can access less than $d^{\bot}$ bits of a codeword, then she gains no information at all on the associated message.

Linear coset-coding for the WT channel can be efficiently implemented using LDPC codes \cite{TDCMM07,STBM10}.

\section{From the Wire-Tap Channel to Non-Malleable Codes}
\label{sec:scheme}

For our construction, we only deal with tampering functions that are bit-wise independent.

\subsection{Motivations for Using Wire-Tap}

Roughly speaking, in both models, codewords are mo\-dified either with random faults (WT I), adversary-controlled erasures (WT II) or an adversary-controlled tampering function (NMC). From these modified codewords or their decoding results, the adversary tries to learn information on the original messages. 

The first WT is a little different from the other models because errors are random and so do not occur in the same number and bit positions every time. It could however be covered by the definition of NMC if every possible tampering caused by these random errors were included in the family of tampering functions taken into account by the code.

Let us now assume that we want to use a linear coset-coding scheme with a parity-check matrix $H$ as NMC. We cannot be protected against tampering functions that only add errors (\ie bit-wise independent functions where the only choices for each bit are \textbf{keep} or \textbf{flip}). To see why, let $\mf$ be a family of such functions. Obviously, for each $f \in \mf$, there is an error vector $e \in \ff_2^n$ such that $\forall c \in \ff_2^n, f(c)=c+e$. Let us follow the tampering experiment. Let $m \in \ff_2^r$ be a source message and $c$ an encoding of $m$. Say $c$ is tampered to $\tilde{c}=c+e$. Decoding results in $\tilde{m}=H ^t \! c + H ^t \! e = m + H ^t \! e$. Thus, $\tilde{m}$ is always $m$ plus a constant offset. It is consequently related to $m$. Linear coset-coding cannot be non-malleable w.r.t. these ``error-only'' functions. There must me some \textbf{0} and \textbf{1} in the tampering.

This is why we consider WT II. Indeed, using \textbf{0} and \textbf{1} on some bits of the codewords is, in an information-theoretic sense, like having erasures at the corresponding locations, as we do not know what was originally there. As WT II guarantees that no information is leaked from erased codewords encoded using an appropriate coset-coding scheme, there will be no relation between the decoded tampered codeword and the original message. That is what motivates our proposal.

\subsection{The Construction}

As discussed before, we consider bit-wise independent functions where the sub-functions are not only \textbf{keep} or \textbf{flip}. Nevertheless, we authorize bit-flips because if the result of the tampering experiment is unrelated to the original message, then the result added to a constant offset will also be unrelated to this message.

We state the following theorem:

\begin{thm}[Linear coset-coding as NMC]
\label{theo}
Let $\mf \subset {\ff_2^n}^{\ff_2^n}$ be a family of bit-wise independent tampering functions such that:

$\forall f=(f_1,\ldots,f_n) \in \mf, |\{i | f_i = \mathbf{0} \ \mathrm{or} \ f_i = \mathbf{1} \} | \geq D$.

Let $C$ be a $[n,k,d]$-linear code such that $D > n-d^{\bot}$, where $d^{\bot}$ is the minimal distance of its dual code $C^{\bot}$.

Then a linear coset-coding using $C$ is non-malleable w.r.t. $\mf$.
\end{thm}

\subsection{Proof of Non-Malleability}

Our proof of non-malleability is inspired from the proof of security of the WT II in \cite{Zem00}.

Let us consider we are in the situation of Theorem~\ref{theo}. Let $f =(f_1,\ldots,f_n) \in \mf$ be a tampering function. Let $S_{\textbf{01}}$ be the set of all positions $i$ such that $f_i = \mathbf{0}$ or $f_i=\mathbf{1}$. Let $S_{\mathbf{keep}}$ and $S_{\mathbf{flip}}$ be the equivalent sets for \textbf{keep} and \textbf{flip}. Let $e \in \ff_2^n$ be such that $\forall i=1,\ldots,n,  e_i=\chi_{S_{\mathbf{flip}}} (i)$ (where $\chi_A$ denotes the indicator function of a set $A$) and $\epsilon \in \ff_2^n$ be such that $\epsilon_i=1$ if $f_i=\mathbf{1}$ and $\epsilon_i = 0$ otherwise. Let $h_1,...,h_n$ denote the columns of the parity-check matrix $H$.

Let $m \in \ff_2^r$ be a message encoded to $c \in \ff_2^n$. Let $\tilde{c}=f(c)$ and $\tilde{m}=H \tilde{c}$. We have 

\begin{eqnarray*}
\tilde{m} &=& \sum\limits_{i \in S_{\textbf{01}}} h_i \tilde{c}_i + 
\sum\limits_{i \in S_{\textbf{keep}}} h_i \tilde{c}_i + \sum\limits_{i \in S_{\textbf{flip}}} h_i \tilde{c}_i \\
 &=& \sum\limits_{i \in S_{\textbf{01}}} h_i \epsilon_i + \sum\limits_{i \in S_{\textbf{keep}}} h_i c_i + \sum\limits_{i \in S_{\textbf{flip}}} h_i (c_i+e_i) \\
&=& H ^t \! \epsilon + H ^t \! e + \sum\limits_{i \in S_{\textbf{keep}} \cup S_{\textbf{flip}}} h_i c_i\\
(\hspace{-0.35cm}&=& m + H ^t \! \epsilon + H ^t \! e - \sum\limits_{i \in S_{\textbf{01}}} h_i c_i )
\end{eqnarray*}

If we want $\tilde{m}$ to be unrelated to $m$, then we want $\sum\limits_{i \in S_{\textbf{keep}} \cup S_{\textbf{flip}}}h_i c_i$ to be unrelated to $m$. If the submatrix $H_{\mathbf{kf}}$ made of the columns $h_i$, $i \in S_{\textbf{keep}} \cup S_{\textbf{flip}}$ is of full rank $r=n-k$, then we gain no information on the corresponding bits of $m$, and all values are equiprobable. This is achieved in particular if $|S_{\textbf{keep}} \cup S_{\textbf{flip}}|<d^{\bot}$ (see chapter 9 of \cite{Zem00}). 

If $D>n-d^{\bot}$, then $|S_{\mathbf{01}}| > n-d^{\bot}$, \ie $n -|S_{\textbf{keep}} \cup S_{\textbf{flip}}| > n-d^{\bot}$ or $|S_{\textbf{keep}} \cup S_{\textbf{flip}}|<d^{\bot}$. The condition of the previous paragraph is thus achieved if we use the parameters of Theorem~\ref{theo}.

Let us define more formally the distribution $D_f$ associated to $f$. Let $K_i$, $i\in S_{\textbf{keep}} \cup S_{\textbf{flip}}$ be Bernoulli(1/2) distributions. Then $D_f = H ^t \! \epsilon + H ^t \! e + \sum\limits_{i \in S_{\textbf{keep}} \cup S_{\textbf{flip}}} h_i K_i$. This distribution and the result of the tampering experiment are identically distributed.

The coset-coding scheme used in Theorem~\ref{theo} is consequently non-malleable w.r.t. $\mf$.

\qed

\subsection{Going Further}

\subsubsection*{Towards a Larger Family of Tampering Functions}

When comparing our construction to the one of \cite{DPW10}, one can relate the LECSS and our coset-coding scheme. The only requirement that is not fulfilled by linear coset-coding is a large distance. As the distance of linear coset-coding is 1, we cannot assume $d > n/4$ as they do. 
That is why we cannot directly modify this construction and replace LECSS with coset-coding in the description of the code and the proof of non-malleability.

Both LECSS and coset-coding ensure non-malleability when the number of $\mathbf{0}$ or $\mathbf{1}$ sub-functions of the tampering function is high enough. To deal with the case where the number of such functions is low, Dziembowski \etal concatenated the LECSS with an AMD code. In such a case, the tampering function acts by adding an error following a fixed distribution (\ie independent of the codeword) and the decoding procedure results in $\bot$ with high probability because of the AMD code. Therefore, non-malleability is ensured. Following this idea, it might also be possible to encapsulate our coset-coding scheme within an error-detecting or an error-correcting code. Thus we would achieve non-malleability w.r.t. a larger family of functions. In particular, functions with a small number of $\mathbf{0}$ or $\mathbf{1}$ sub-functions which cannot be dealt with by coset-coding alone could be included. For the error-detecting case, using an AMD code as in \cite{DPW10} seems to be feasible. However, for the error-correcting case, it is not clear which kind of correction strategy to use to deal with the effects of the linear coset-coding scheme.
Nevertheless, if such functions are the only ones of interest, one must be aware that an error correcting or an error detecting code is sufficient by itself. 

\subsubsection*{Relaxing the Notion of Non-Malleability}

In the model for the WT II described in this paper, we require that Eve cannot obtain any bit of information on the messages sent over the channel. This strong security notion can be relaxed. Indeed, one could be satisfied even if Eve learned only a bounded amount of bits. This is possible if we consider generalized Hamming distances \cite{Wei91} instead of the dual distance $d^{\bot}$ of the code considered in the linear coset-coding scheme. For $i \in \mathbb{N}$, the generalized distance $d_i$ is such that if Eve cannot obtain more than $d_i$ bits per message, then she gains no more than $i-1$ bits of information per message. For instance, $d_1 = d^{\bot}$.

In the same spirit, one could relax the notion of non-malleability. After the tampering experiment, we could state that either the decoding procedure returns the original message or it enables to learn a bounded number of bits of information on this message. Using our construction, it is easy to build another scheme that would satisfy this requirement. One would only have to replace dual distances by generalized distances.

\section{Conclusion}
\label{sec:conclu}

We established in this paper a parallel between Non-Malleable Codes and the Wire-Tap Channel. This relation enabled us to build an efficient non-malleable scheme,  w.r.t. a family of bit-wise independent functions, that is neither error-correcting nor error-detecting.

Considering bit-wise independent tampering is a worthwhile first step for NMC. An interesting open problem would be now to build schemes that are non-malleable w.r.t. larger families of functions.

\section*{Acknowledgement}

The authors would like to thank Julien Bringer for his helpful comments.

\bibliographystyle{IEEEtran}
\bibliography{IEEEabrv,biblio}

\end{document}